\documentclass[aps,prb,twocolumn,showpacs]{revtex4}
\usepackage[dvips]{graphicx}
\usepackage{amsmath}
\usepackage{times}

\begin{document}
\author{ %
D. V. Savin and H.-J. Sommers}
\affiliation{ %
Fachbereich Physik, Universit\"at Duisburg-Essen, 45117 Essen, Germany}

\title{Shot noise in chaotic cavities with an arbitrary number of open channels}
\received{23 December 2005} %
\published{13 February 2006 in: %
\underline{Phys. Rev. B \textbf{73}, 081307(R) (2006)} }

\begin{abstract}
Using the random matrix approach, we calculate analytically the average
shot-noise power in a chaotic cavity at an arbitrary number of propagating
modes (channels) in each of the two attached leads. A simple relationship
between this quantity, the average conductance and the conductance variance
is found. The dependence of the Fano factor on the channel number is
considered in detail.
\end{abstract}


\maketitle

The time dependent fluctuations in electrical currents caused by random
transport of the electron charge $e$, which (unlike thermal fluctuations)
persist down to zero temperature, are known as shot noise. In mesoscopic
systems, an adequate description of this phenomenon is achieved in the
scattering theory framework.\cite{Blanter2000,Beenakker1997} In particular,
for the two-terminal setup (with a small voltage difference $V$) it is
well-known that the zero-frequency shot-noise spectral power is given by
\cite{Khlus1987,Lesovik1987,Buettiker1990}
\begin{equation}\label{Pdef}
P = P_0 \sum_{p=1}^{n}T_p(1-T_p)\,,\quad P_0=2e|V|G_0\,,
\end{equation}
where $T_p$ are $n=\mathrm{min}(N_1,N_2)$ transmission eigenvalues of a
conductor, $G_0$ is the conductance quantum, and $N_{1,2}$ denotes the
number of scattering channels in each of the two leads. $T_p$ are mutually
correlated random numbers between 0 and 1 whose distribution depends on the
type of the conductor.

In the case of chaotic cavities considered below, universal fluctuations of
$T_p$ are believed to be provided by the random matrix theory
(RMT).\cite{Baranger1994,Jalabert1994} The latter is characterized by the
symmetry index $\beta$, distinguishing between universality classes of
systems according to the absence ($\beta=2$, unitary ensemble) or presence
($\beta=1$, orthogonal ensemble) of time-reversal symmetry and spin-flip
symmetry ($\beta=4$, symplectic ensemble). Various RMT related aspects of
the shot noise are under active study now, both theoretically
\cite{Blanter2000i,Agam2000,Nazmitdinov2002i,Silvestrov2003i,Jacquod2004,Sukhorukov2005,
Aigner2005} and experimentally\cite{Oberholzer2001,Oberholzer2002} (see also
the references in these papers). However, exact results for the average
shot-noise power $\langle{P}\rangle$ were reported in the literature only in
the limiting cases of $N_{1,2}\gg1$\cite{Beenakker1997,Nazarov1995} (which
is the purely classical one)\cite{Blanter2000i,vanLangen1997} or
$N_1=N_2=1$,\cite{Pedersen1998} the experimentally relevant case of few
channels being an open problem.

An alternative consideration was undertaken very recently by Braun et
al.,\cite{Braun2005} who developed the semiclassical trajectory approach to
build up the $1/N$ expansion for $\langle{P}\rangle$, extending earlier
results\cite{Schanz2003i,Richter2002} to all orders of the inverse total
number of channels, $N=N_1+N_2$ (see also Ref.~23). They were able (for
$\beta=1,2$) to sum up the resulting series in a compact form, which we
represent introducing $\beta$ as follows:
\begin{equation}\label{P}
\frac{\langle{P}\rangle}{P_0} =
\frac{N_1(N_1-1+\frac{2}{\beta})N_2(N_2-1+\frac{2}{\beta}) }{
(N-2+\frac{2}{\beta})(N-1+\frac{2}{\beta})(N-1+\frac{4}{\beta})}\,.
\end{equation}
This result surprisingly turned out to remain valid down to $N_{1,2}=1$, as was
checked by comparison to numerics.

Our aim here is to provide the exact RMT derivation of Eq.~(\ref{P}) valid
at arbitrary $N_{1,2}$ and all $\beta$. There are several ways to perform
the calculation. First, $T_{p}$ are defined as the singular values of a
transmission matrix $t$ (which is a $N_1{\times}N_2$ off-diagonal block of a
$N{\times}N$ unitary scattering matrix).\cite{Buettiker1990} Finding
$\langle{P}\rangle=P_0\langle\mathrm{tr\,}[tt^{\dagger}(1-tt^{\dagger})]\rangle$
amounts thus to an integration over the unitary group which is a quite
complicated problem in general.\cite{Brouwer1996} Second, one can think of
(\ref{Pdef}) as a linear statistic on the transmission eigenvalues, so that
$\langle{P}\rangle=P_0\int_0^1dT\rho(T)T(1-T)$ is provided by the
transmission eigenvalue density $\rho(T)$. Unfortunately, the latter is
explicitly known only in the above-mentioned limiting cases.

We follow below yet another route. Contrary to the density $\rho(T)$, the joint
probability distribution function $\mathcal{P}_{\beta}(\{T_p\})$ of all
transmission eigenvalues is known\cite{Beenakker1997} to have the following
attractively simple form at arbitrary $N_{1,2}$:
\begin{equation}\label{Joint}
\mathcal{P}_{\beta}(\{T_p\}) = \mathcal{N}_{\beta}^{-1} |\Delta(T)|^{\beta}
\prod_{j=1}^{n} T_j^{(\beta/2)(|N_2-N_1|+1)-1}\,,
\end{equation}
where $\Delta(T)=\prod_{i<j}(T_i-T_j)$ is the Vandermonde determinant. The
key idea is to appreciate a relation of (\ref{Joint}) to the (integral
kernel of) Selberg's integral defined as follows:\cite{Mehta2}
\begin{eqnarray}\label{Selberg}
I(a,b,c,n) &\equiv& \int_{0}^{1}\!\!\cdots\!\!\int_{0}^{1} |\Delta(T)|^{2c}
\prod_{j=1}^{n} T_j^{a-1}(1-T_j)^{b-1} dT_j \nonumber\\
&=& \prod_{j=0}^{n-1}\frac{\Gamma(1+c+jc)\Gamma(a+jc)\Gamma(b+jc)}{
\Gamma(1+c)\Gamma(a+b+(n+j-1)c)} ,
\end{eqnarray}
with $\Gamma(x)$ being the gamma function. This result [as well as Eqs.
(\ref{rec1}) and (\ref{rec2}) below] holds generally for complex $a$, $b$
and $c$ with positive real parts.\cite{remark} One readily sees that
(\ref{Joint}) corresponds to the following particular values of these
parameters
\begin{equation}\label{abc}
a=(\beta/2)(|N_2-N_1|+1)\,,\quad b=1\,,\quad\mathrm{and}\quad c=\beta/2\,.
\end{equation}
It is worth noting that at these  values the second line of (\ref{Selberg})
provides us with the normalization constant $\mathcal{N}_{\beta}$.

Selberg's integral can be seen as a multidimensional generalization of
Euler's beta function. Due to the specific structure of the integral kernel
in (\ref{Selberg}) very useful recursion relations may be established for
certain moments (see Ref.~25). In particular,
\begin{equation}\label{rec1}
\langle{T_1^2}\rangle =
\frac{[a+1+2c(n-1)]\langle{T_1}\rangle-c(n-1)\langle{T_1T_2}\rangle}{
a+b+1+2c(n-1)}
\end{equation}
and
\begin{equation}\label{rec2}
\langle{T_1T_2\cdots T_m}\rangle = \prod_{j=1}^m
\frac{a+c(n-j)}{a+b+c(2n-j-1)}
\end{equation}
are relevant for the problem discussed. Putting above $m=1$ and taking into
account (\ref{abc}), one easily gets from the Landauer formula the
average conductance as follows:
\begin{equation}\label{G}
\langle{G}\rangle = G_0n\langle{T_1}\rangle =
G_0\frac{N_1N_2}{N-1+2/\beta}\,,
\end{equation}
in agreement with Baranger and Mello.\cite{Baranger1994} Equation
(\ref{rec2}) at $m=2$ gives
$\langle{T_1T_2}\rangle=\langle{T_1}\rangle(\mathrm{max}(N_1,N_2)-1)/(N-2+\frac{2}{\beta})$
and then $\langle{T_1^2}\rangle$ follows from (\ref{rec1}) that allows us
to obtain after some simple algebra the following result for the conductance
variance,
$\mathrm{var}(G/G_0)=n\langle{T_1^2}\rangle+n(n-1)\langle{T_1T_2}\rangle
-n^2\langle{T_1}\rangle^2$:
\begin{equation}\label{varG}
\frac{\mathrm{var}(G)}{G_0^2} =
\frac{2N_1(N_1-1+\frac{2}{\beta})N_2(N_2-1+\frac{2}{\beta}) }{
\beta(N-2+\frac{2}{\beta})(N-1+\frac{2}{\beta})^2(N-1+\frac{4}{\beta})}\,.
\end{equation}
This result was derived earlier\cite{Beenakker1997,Baranger1994,Jalabert1994,Iida1990}
by a different method. Finally, along the same lines we arrive at expression
(\ref{P}) for $\langle{P}\rangle=P_0n(\langle{T_1}\rangle-\langle{T_1^2}\rangle)$ .

Comparing the average shot-noise power (\ref{P}), conductance (\ref{G}), and
conductance variance (\ref{varG}), one immediately finds the following
relationship between them at arbitrary $N_{1,2}$:
\begin{equation}\label{PvarG}
\frac{2}{\beta} \frac{G_0}{P_0}
\frac{\langle{P}\rangle\langle{G}\rangle}{\mathrm{var}(G)} =  N_1N_2 \,.
\end{equation}
It would be interesting to understand whether such a relation holds for
other types of mesoscopic conductors and how it is modified in other regimes
(e.g., in the crossover between ensembles).

\begin{figure}[t]
\includegraphics[width=0.425\textwidth]{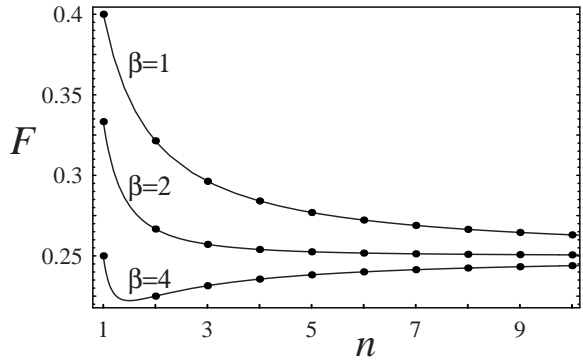}
\caption{The Fano factor as a function of the channel number in symmetric
($N_1=N_2=n$) chaotic cavities of different RMT ensembles.}
\end{figure}

We proceed now with the discussion of the obtained results. In the case of
uncorrelated electrons, electron transfer is a Poisson process that results
in the value $P_P=2e\langle{I}\rangle=P_0\langle{G}\rangle$ for the mean
power. The suppression of the actual noise (\ref{Pdef}) with respect to this
Poisson value is customarily described by the Fano factor
$F=\langle{P}\rangle/P_P$. One finds from (\ref{P}) and (\ref{G}) that
\begin{equation}\label{F}
F  = \frac{(N_1-1+2/\beta)(N_2-1+2/\beta) }{ (N-2+2/\beta)(N-1+4/\beta)}
\end{equation}
at arbitrary $N_{1,2}$. In the semiclassical limit of large number of
channels, $N_{1,2}\gg1$, one readily gets from (\ref{F}) $F\approx
N_1N_2/N^2-(1-\frac{2}{\beta})(N_1^2-N_1N_2+N_2^2)/N^3$, i.e. the known
classical value and the first weak-localization
correction.\cite{Beenakker1997,Jalabert1994}

In the symmetric case, $N_1=N_2=n$, Eq.~(\ref{F}) reduces to
\begin{equation}\label{FN}
F = \frac{(n-1+2/\beta)^2}{ (2n-2+2/\beta)(2n-1+4/\beta)}\,.
\end{equation}
The Fano factor starts from the value $\frac{2}{\beta+4}$ ($=\frac{2}{5}$,
$\frac{1}{3}$, $\frac{1}{4}$ for $\beta{=}1$, 2, 4, respectively) at $n=1$
and tends to the classical value $\frac{1}{4}$ as $n\to\infty$. For the
orthogonal or unitary ensemble ($\beta=$1 or 2) this is a monotonic decrease
in $n$, whereas for symplectic ensemble ($\beta=4$) $F$ has a minimum
$\approx0.225$ at $n\approx2$. Figure 1 illustrates these dependencies. The
shot noise is always suppressed more strongly in the symplectic case.

\begin{figure}[b]
\includegraphics[width=0.45\textwidth]{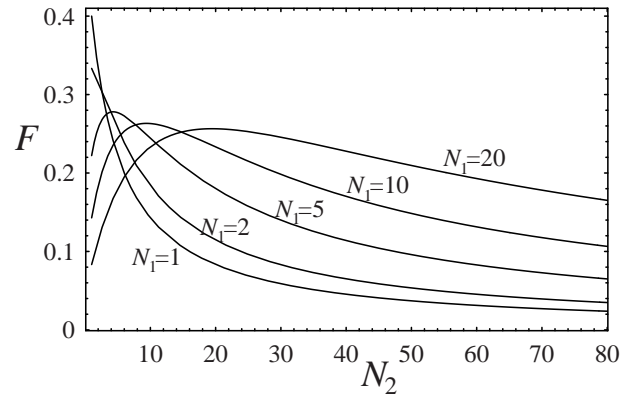}
\caption{The Fano factor at fixed number $N_1$ of channels in one lead
and varied one $N_2$ in the other lead. The value at the maximum at
$N_2\approx N_1$ is close to
$F_\mathrm{max}^{*}\approx\frac{1}{4}-\frac{1}{8}(1-\frac{2}{\beta})N_1^{-1}$
if $N_1\gg1$. Plotted is the result for cavities with time-reversal symmetry
($\beta=1$).}
\end{figure}

In the general case of asymmetric cavities, it is instructive to consider
the Fano factor at given fixed number $N_1$ of channels in one lead as a
function of the channel number $N_2$ in the other lead. One easily finds
from (\ref{F}) (see Fig.~2) that $F$ starts from the value  $\frac{2}{\beta
N_1+4}$ at $N_2=1$ and then develops a maximum at
$N_2^*=\sqrt{(N_1-1)(N_1+2/\beta)}+1-2/\beta$, taking the following value at
the maximum:
\begin{equation}
F_\mathrm{max}^{*}=\frac{N_1+2/\beta-1}{2\sqrt{(N_1-1)(N_1+2/\beta)}+2N_1+2/\beta-1}\,.
\end{equation}
As $N_2$ grows further, $F$ decreases down to zero according to
$F\approx(N_1-1+\frac{2}{\beta})/N_2$. This fact could be understood
qualitatively: the lead with $N_2\gg1$ becomes almost classical with a
deterministic transport through it that suppresses fluctuations of $T_p$,
thus $P\to0$.\cite{Beenakker1991} Such a suppression of the shot noise in
strongly asymmetric cavities was indeed observed in the recent
expreriment.\cite{Oberholzer2001} [We note, however, that this experiment
deals with asymmetric cavities when both $N_{1,2}$ are large, their ratio
$\eta=N_2/N_1$ being varied. In this case the classical result
$F=\eta/(1+\eta)^2$ applies.]

In summary, we have exactly calculated the average shot noise power at an
arbitrary number of open channels by relating the problem to Selberg's
integral. The proposed method is not restricted by linear statistics only
and may be applied further to study, e.g., higher-order charge fluctuations
as well as the whole distribution of shot-noise power. It would be highly
interesting to check experimentally the predicted finite $N$ behavior
(\ref{F}) of the Fano factor.

Recently, we became aware of the related study by Bulgakov et
al.\cite{Bulgakov2006} done at $N_1=N_2$ and $\beta=1,2$. We thank V. Gopar
for this communication.

We are grateful to P.~Braun, F.~Haake, S.~Heusler and S.~M\"uller for useful
discussions. The financial support by the SFB/TR 12 of the DFG is
acknowledged with thanks.


\begin{thebibliography}{26}
\bibitem{remark}
  Actually, one has even a weaker restriction on $c$:\\
  $\mathrm{Re\,}c>-\mathrm{min}[1/n,\mathrm{Re\,}a/(n-1),\mathrm{Re\,}b/(n-1)]$.
\end{thebibliography}

\end{document}